\documentclass[preprint,prb]{revtex4}
\usepackage{graphicx}
\begin{document}

\title{Photo-induced volume changes in selenium.
 Tight-binding molecular dynamics study.}
\author{J. Heged\"us$^1$,  K. Kohary$^{1,2}$, S. Kugler$^1$, 
and K. Shimakawa$^3$}
\affiliation{$^1$Department of Theoretical Physics,
Budapest University of Technology and Economics,
H-1521 Budapest, Hungary,}
\affiliation{$^2$Department of  Materials, University of Oxford,
Parks Road, Oxford OX1 3PH, UK}
\address{$^3$Department of Electrical and Electronic Engineering,
Gifu University, Gifu 501-1193,
Japan}


\begin{abstract}

Tight-binding molecular  dynamics simulations of  photo-excitations in
small  Se clusters  (isolated Se$_8$  ring  and  helical Se 
chain)  and
glassy Se networks (containing 162 atoms) were carried out in order to
analyse the  photo induced instability  inside the amorphous selenium. 
 In the cluster 
systems after  taking an
electron  from the  highest  occupied  molecular  orbital  to  the  lowest
unoccupied molecular  orbital a bond breaking occurs.
In the glassy networks photoinduced volume expansion was observed and at
the same time the number of coordination defects changed significantly
due to  illumination.



PACS number: 73.61.Jc \hfill \break
Keywords:
 Chalcogenides;   Photoinduced phenomena

\end{abstract}

\maketitle
\section{Introduction}

The   photo-induced   structural   changes  during      illumination
\cite{Shimakawa95,Morigaki99}    in  chalcogenide   glasses   can  be
classified into two groups, those which are irreversible and which can
be reversed.  An other possible sorting is the following: there
are defect related metastability  and there are structure related one.
Several  investigations have  been carried  out  in amorphous selenium 
(a-Se)
  in order  to
provide    an    acceptable    explanation    of    these    phenomena
\cite{Fritzsche96,Kolobov99,Zhang99,Ganjoo00,Palyok01}.  
 Present  capacity of
fast  computers  allow  us  to simulate  the  photo-induced  structural
changes  i.e. to  follow the  time development  of these  phenomena by
molecular dynamics (MD).  Our particular aim was to  determine how the
structural changes  occur in a-Se  due to illumination.   We performed
tight-binding  molecular dynamics simulations  to study the dynamics
 of a  Se$_8$  ring, a linear
chain and glassy Se networks  before, during and
after the illumination.

\section{Simulation method}

 We have  developed a  MD computer code  (ATOMDEP program  package) to
simulate the  real preparation  procedures of   amorphous structures
(growth  by   atom-by-atom  deposition   on  a  substrate   and  rapid
quenching).   In   our  recent  work\cite{Kohary01},   the  growth  of
amorphous  carbon  films  were  simulated  by this  MD  method.   This
computer  code  is  convenient  to  investigate  photo-induced  volume
changes as well  if  the  built-in  atomic  interaction  can  handle  the
photo-excitation.  Standard  velocity Verlet algorithm  was applied in
our MD  simulations in  order to follow  the atomic scale  motion.  We
chose ${\Delta}t$ =  1~fs or 2~fs for the time  step, depending on the
temperature. These  values are the  typically used time steps  for such
simulations.  To  control the temperature  of the systems we  applied
the velocity-rescaling method.

For calculating  the inter-atomic forces in a-Se  we used tight-binding
(TB)   \cite{Molina99}  and  self-consistent   tight-binding  (SCF-TB)
\cite{Lomba00}  models. The TB parametrization \cite{Molina99} has  been
introduced  for  disordered  selenium  following  the 
techniques  developed   by  Goodwin  et  al \cite{Goodwin89}.  It  was
thoroughly  tested by  molecular dynamics  calculations in  liquid and
amorphous phases and the results were
  compared to experiments and to  {\it ab initio}
calculations.   The agreement  with  experiments  and {\it  ab
initio }  calculations are  rather good apart  from the fact  that the
number  of coordination  defects in  the solid  and liquid  phases are
higher than the experimentally  measured values.  The authors improved
their  TB Hamiltonian by including   the  Hubbard
correction   \cite{Lomba00}.  
This implies that either the algorithm have to be made self-consistent
or perturbation theory must be applied. Our choice was 
the first alternative. 
  The   convergence  for   this   SCF-TB
calculations  was controlled by  the charge  differences on  the atomic
sites. 
Convergence criteria were considered to be satisfied
if  the deviation of atomic charges between the
actual and  the previous iterations was  less than 0.01 electron/atom.

\section{Sample preparation}
\subsection{Cluster geometries}

  Crystalline forms  of selenium consist of  chains and eight-membered
  rings. Typical bond lengths are  around 2.35~\AA \ while most of the
  bond angle  values are around $103^{\circ}$. It is very likely
 that these
 local arrangements can be found in non-crystalline forms of selenium
as well.  The  initial  configuration  of the eight-membered  ring  in  our
  simulation  had  bond  lengths of 2.38  \AA \  and  bond  angles  of
  102$^{\circ}$.   Dihedral  angles  were equal to  100$^{\circ}$  (See
  Fig. 1. top panel).  For the eighteen-membered  selenium chain (with
  1-dimensional periodic   boundary  condition)  these   values  were:
   2.36~\AA,  \
  100$^{\circ}$, and 98$^{\circ}$, respectively. Every Se atom had two
  first-neighbors, i.e.  there was no coordination defect.

\subsection{Glassy structure}

Glassy  selenium networks  having  162 atoms  (with three  dimensional
 periodic boundary  conditions) were  prepared by ``cook  and quench''
 technique  (from 3000 K  to 20  K).  The  average quenching  rate was
 $4\cdot10^{12}$ K/s.  In our simulation after  quenching the periodic
 boundary  constraint in  the z  direction  was released  and and  the
 system was  allowed to  relax for  another 100 ps.   In this  way the
 system could change  its volume in the z  direction. Four models have
 been prepared  with different initial topologies  and densities.  The
 snapshot of one of our models can be seen in Fig.2.

\section{Results}

\subsection{Photo excitation of small Se clusters}

   Before illumination (photo excitation)  the individual ring and the
chain  were relaxed  for 4  ps  at T=500  K.  During  this period  the
structures were  stable.  When a  photon is absorbed an  electron from
the highest occupied molecular orbital (HOMO) to the lowest unoccupied
molecular orbital  (LUMO) is transferred.   This is a simple  model of
photo-excitation when an electron is  shifted from the valence band to
the  conduction  band (electron-hole  pair  creation).   In our  study
electron-hole pair recombination was also simulated by taking back the
electron from the  conduction  band to the valence band. 
  The lifetime of the excitation  was 200 fs. 

  After excitation one bond length in the ring started to increase and
bond-breaking occurred.  Two  snapshots of this process can  be seen in
 Figure 1. A  similar result  was published  \cite{Shimojo98}  by a
Japanese  group for  S$_8$. They  performed MD  simulation  within the
framework  of   density  functional   theory  in  the   local  density
approximation.
In  our  second  MD   simulation  we  investigated  the  linear  chain
structure. The same procedure was  performed to model the excitation.
  An electron
from HOMO to  LUMO was transferred. Very similar  result was obtained;
 a
bond inside  the chain  was broken immediately  after a HOMO 
 electron was
excited.

\subsection{Photo excitation of glassy Se networks}

The excitation procedure for the glassy networks was different from the
method applied for small clusters. The hole procedure was carried out at the 
temperature of 20 K.
Instead of a single electron excitation, five
electrons were transferred from the valence band to the
conduction band.
 The first  excited
electron  was taken  from the  HOMO, the  second excited  electron was
taken from the level below  it and so on.
A difficulty applying this procedure  appeared during the  simulation.
The  HOMO
and the LUMO  levels shifted into the  middle of the  gap after excitation
and even level crossing was observed. After that point the  result
obtained  by MD  simulation was  uncertain and  useless. It could cause
incorrect conclusions.
 To  avoid this
problem we  applied the  following calculation method:
  we separately
calculated the influences of  electron and hole creation and 
annihilation. 

 Our systems were open only along the z direction 
and periodic boundary conditions were applied in the
other two directions.
To measure the length (i.e. volume) of the sample we 
have proceeded in the following
way.   We have considered
40 atoms in z direction closest to the either ends of the  sample.
But we neglected those 10-10 atoms at the ends of these
clusters, which had the largest (smallest) z coordinated to
avoid the surface effects.
    The length of the  sample was
defined as the average z coordinates  of 30 rest atoms at the left end
of the sample minus the average  z coordinates of 30 rest atoms at the
right end.

The total length of four samples with
 average density of 4.5 g/cm$^3$ is  shown in Fig. 3. 
Arrows demonstrate the electron excitations
and  deexcitations.  Only one  excitation  out  of  five caused  volume
decrease.  The  total influence is  volume expansion. In  the opposite
process three out of five  decrease, one increases the length.  Final
conclusion  is   that  photo-induced  volume expansion occurred during 
illumination and remained
 after
finishing  illumination.  A typical  atomic scale  rearrangement after
one  photo-induced  electron  excitation  is  shown in  Fig.  5. 

  More
interesting is  the number of coordination  defects. Basically, selenium
atoms  have two  nearest-neighbors.  
Average numbers of one-fold coordinated and threefold-coordinated atoms
can  be  seen  in  Fig  4. (Practically,  the number  of twofold
coordinated atoms remained the same in our 
simulations.) During   illumination  the  number  of
one-coordinated  defects  was increasing  while  it  was opposite  for
threefold coordinated  selenium atoms. During the  deexcitation  the
inverse process occurred. It is  striking that there are more threefold
defects and  less one-fold  coordinated atomic sites  at the end  of the
simulation compared to initial configuration.
 All these observations suggest the atomistic
explanation of volume changes in chalcogenide systems.

\section{Summary}

  We have developed a  molecular dynamics computer code to investigate
the  time   development  of  ring  and   chain-like  structures  after
photo-excitation. Furthermore, we investigated the photo-induced volume
change  in   glassy  networks.   Our   results  suggest
photo-induced bond  breaking and the  change of the number  in  coordination
defects  in  these  conditions. This  might  be  an  explanation  of
photo-induced volume changes in a-Se and other chalcogenide glasses.

\section{Acknowledgments} 

This work has been supported by the OTKA Fund (Grant No.  T043231) and
by Hungarian-British and  Hungarian-Japanese intergovernmental S and T
Programmes   (No.  GB-17/2003   and  No.   JAP-7/2000).   The  computer
simulations  were  partly  carried   out  at  the  Tokyo  Polytechnics
University (Japan). We are indebted to Prof. T. Aoki  (TPU) for providing
us this possibility.


\newpage
\begin{figure}
\caption{
Two snapshots of bond-breaking process caused by photo-excitation
in the  Se$_8$ ring.
Top panel displays the ring structure before excitation.
An electron from the HOMO is excited to the LUMO at t=0 fs. Bottom
panel shows the configuration after 420 fs.
}
\begin{center}
\vspace {1cm}
\vspace{1cm}
\fbox{\parbox{5cm}{ -10\ fs \includegraphics[width=3cm]{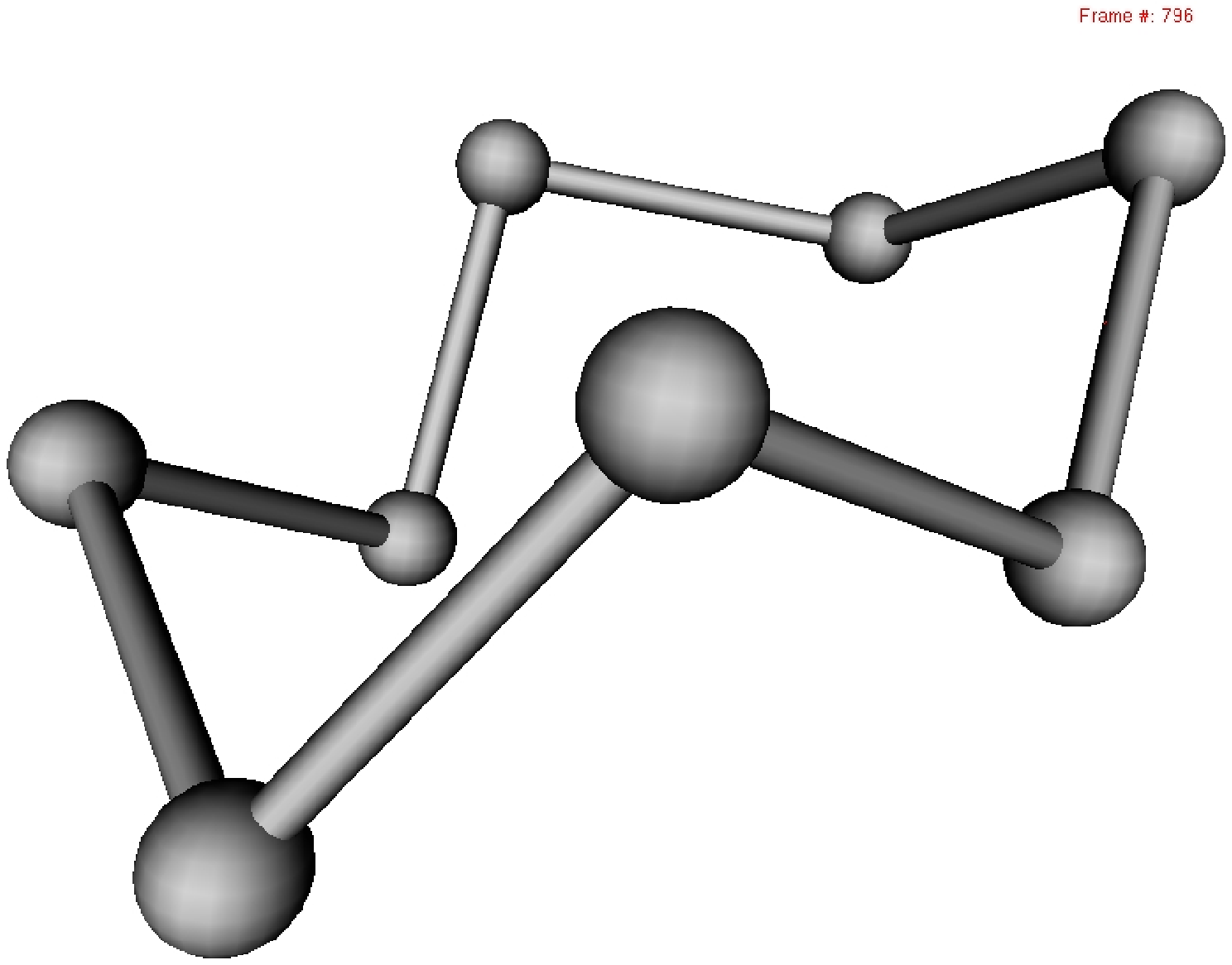}}}\\
\fbox{\parbox{5cm}{ 420 fs \includegraphics[width=3cm]{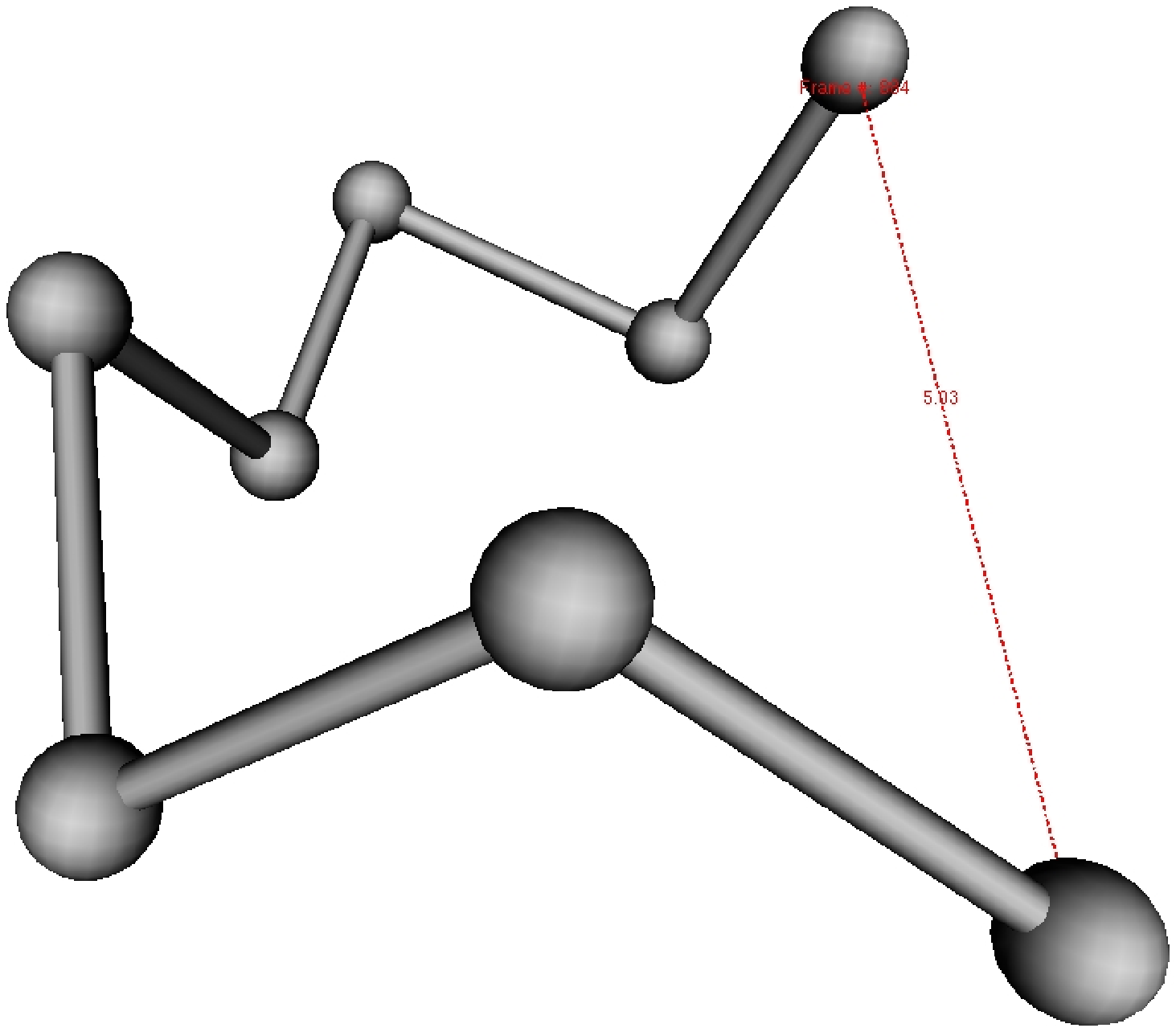}}}
\end{center}
\end{figure}

\begin{figure}
\caption{One of the glassy networks constructed with rapid cooling technique. It contains 162 Se atoms with periodic boundary condition in x and y directions.
}
\begin{center}
\vspace {1cm}
\vspace{1cm}
\includegraphics[width=10cm]{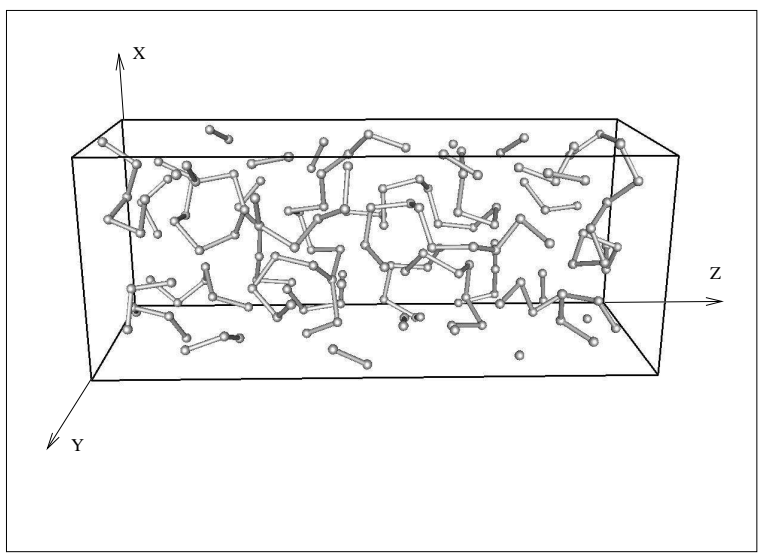}
\end{center}
\end{figure}

\begin{figure}
\caption{
 Total length of four samples with average densities of 4.5 
g/cm$^3$.
The arrows represent the electron excitations (up arrow) and recombinations (down arrow) during the simulations.
}

\begin{center}
\vspace {1cm}
\vspace{1cm}
	\includegraphics[width=9cm]{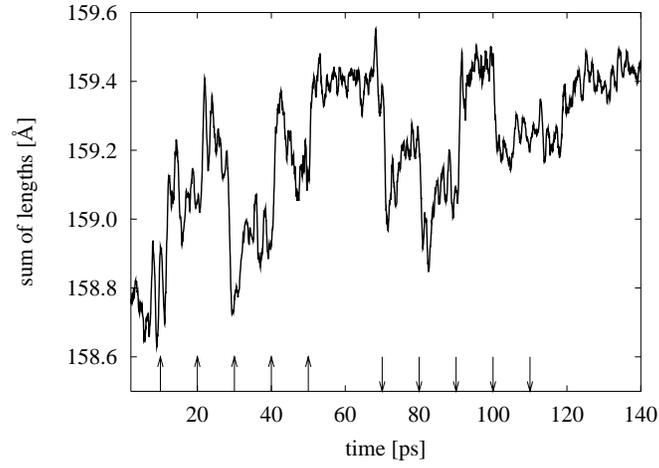}\\
\end{center}
\end{figure}

\begin{figure}
\caption{
Bond breakings and topological rearrangement inside the cluster
 due to illumination in a-Se.
}

\begin{center}
\vspace {1cm}
\vspace{1cm}
	\includegraphics[width=9cm]{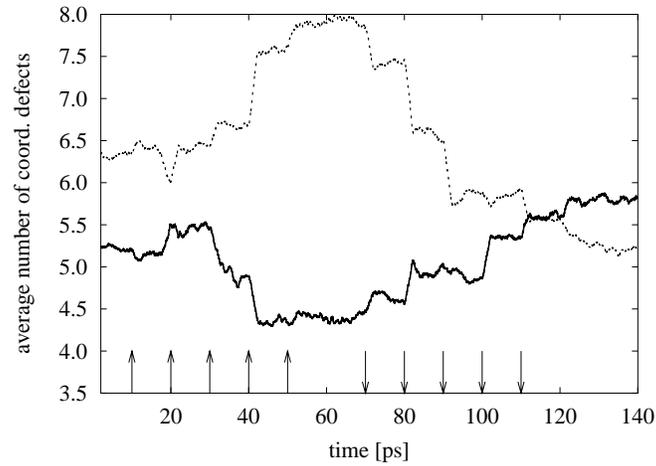}\\
\end{center}
\end{figure}

\end{document}